\documentstyle[numreferences,proceedings,psfig,amssymb,citesort]{crckapb}
\begin{opening}
\title{Is there a critical acceleration for the onset of convection?}
\author{Thorsten P{oe}schel and Thomas Schwager}
\institute{Humboldt-Universit\"at zu Berlin,
Institut f\"ur Physik, \\
Invalidenstra\ss e 110,
D-10115 Berlin, Germany}
\end{opening}

\runningtitle{critical acceleration for the onset of convection?}

\begin{document}
\footnotetext{\bf In: H.J.Herrmann, J.-P. Hovi, and S.Luding (eds): ``Physics of Dry Granular Materials'', Kluwer (Dortrecht, 1998), p.625}
\begin{abstract}
  Suppose granular material is shaken vertically with
  $z(t)=A_0\cos(\omega_0 t)$. Can we expect to find convection if
  $A_0\omega_0^2 < g$? By means of theoretical analysis and computer
  simulation we find that there is no critical
  $\Gamma=\left|A_0\right|\omega_0^2/g$ for the onset of convection.
  Instead we propose a modified criterion which coincides with
  $\Gamma=1$ for small frequency $\omega_0$.
\end{abstract}
\section{Introduction}
In vertically shaken granular material one observes various
macroscopic effects, such as
convection~\cite{EhrichsJaegerKarczmarKnightKupermanNagel:1994,KonvMD,TaguchiPRL},
surface
fluidization~\cite{WarrHuntleyJacques:1995,EvesqueSzmatulaDenis:1990,LudingHerrmannBlumen:1994},
spontaneous heap
formation~\cite{Faraday:1831,DouadyFauveLaroche:1989}, surface
patterns~\cite{MeloUmbanhowarSwinney:1994,MetcalfKnightJaeger:1997,AokiAkiyama:1996,ClementVanelRajchenbachDuran:1996},
oscillons~\cite{MeloUmbanhowarSwinney:1996} and others. The common
feature of all these effects is that particles change their position
with respect to each other. If a single hard particle is shaken, one
agrees that it cannot move as long as during the entire period of
shaking the acceleration due to gravity $g$ exceeds the upwards
acceleration of the shaker. For the case of sinusoidal motion
$z(t)=A_0\cos(\omega_0 t)$ with amplitude this condition is given if
$\Gamma=A_0\omega_0^2/g < 1$.

We want to discuss the problem whether in a granular material
particles at the surface can separate from each other even if
$\Gamma<1$. At first glance the problem seems to be trivial, however,
one has to take into account that there is a {\em nonlinear}
interaction of touching grains.

There is a controversial discussion in the literature whether there is
a critical value of Froude number
$\Gamma_c=\left|A_0\right|\omega_0^2/g$ below which the above
mentioned effects vanish. In {\em many} experimental
observations~(e.g.~\cite{EhrichsJaegerKarczmarKnightKupermanNagel:1994,EvesqueSzmatulaDenis:1990,DouadyFauveLaroche:1989,MeloUmbanhowarSwinney:1994,NurOberhalbG,Moreau:1993})
and computer simulations (e.g.~\cite{Moreau:1993}) such a critical
number $\Gamma_c$ was found. Several authors believe that the value is
$\Gamma_c=1$. In numerical simulations, however, some authors have
found surface fluidization and convection for $\Gamma \lesssim
1$~\cite{TaguchiPRL,LudingHerrmannBlumen:1994,BarkerMehta:1992}.
Gallas, Herrmann and
Soko{\l}owski~\cite{GallasHerrmannSokolowski:1992FLUI} concluded
already 1992 from their numerical results ``\dots that the current
belief that $\Gamma$ determines the degree of fluidization is
incorrect''.

\section{A one dimensional continuum model}
We discuss the response of granular material with respect to vertical
oscillation in the limit of a one dimensional approach: the lowest
bead of a vertical column of $N$ identical spherical beads is shaken
with $z_0=A_0\cos\omega_0 t$ and the other beads move due to their
interaction force and gravity $g$. Adjacent spheres interact with
their next neighbors by the force
\begin{equation}
  \label{forcelaw}
  F_{k,k+1}=-\sqrt{R}\left(\mu\xi_{k,k+1}^{3/2}+\alpha\dot{\xi}_{k,k+1}
\sqrt{\xi_{k,k+1}}\right)~~,
\end{equation}
with $\mu$ and $\alpha$ being elastic and dissipative material
constants, i.e. functions of Young modulus, Poisson ratio and
dissipation rate (for details see~\cite{BSHP}). $\xi$ is the overlap
$2R-\left|z_k-z_{k+1}\right|$ of adjacent spheres of radius $R$ and
mass $m$ at positions $z_k$, $z_{k+1}$. The height of the column is
$L=2NR$. The net force experienced by the $k$-th bead reads, therefore
\begin{eqnarray}
  && F_{k,k+1}-F_{k-1,k}=m\ddot{z}_k = \nonumber\\
  && ~~~~\sqrt{R}\mu\left[\xi_{k-1,k}^{3/2}-\xi_{k,k+1}^{3/2}\right] + 
  \sqrt{R}\alpha\left[\dot{\xi}_{k-1,k}\sqrt{\xi_{k-1,k}} 
  -\dot{\xi}_{k,k+1}\sqrt{\xi_{k,k+1}}\right]~~
  \label{netforce}
\end{eqnarray}
Introducing new coordinates $u_k=z_k - 2Rk$ ($k=0\dots N$) the overlap
of two adjacent spheres can be written as
\begin{eqnarray}
  \xi_{k,k+1}&=&2R-z_{k+1}+z_k = u_{k}-u_{k+1} = u(2Rk) 
  - u(2R(k+1))\label{overlapa}\\
  &=& u(2Rk)-u(2Rk)- 2R\left.\frac{\partial u}{\partial z}\right|_{z=2Rk} = 
  -2R\frac{\partial }{\partial z}u_k\,.
\end{eqnarray}
In (\ref{overlapa}) the discrete value $u_k$ has been replaced by
$u(2Rk)$ which allows for a Taylor expansion. Eq.~(\ref{netforce})
turns into
\begin{eqnarray}
    m \ddot{u}_k &=& \frac{4}{3}\pi R^3\rho\ddot{u}_k = - \mu \sqrt{R} 
   (2 R)^\frac{3}{2} A -\alpha\sqrt{R}\left(2R\right)^\frac32 B\label{eqnofmotion1}\\
\ddot{u}_k &=& -\frac{3}{\sqrt{2}\pi\rho R}\left(A\mu+B\alpha\right)
\end{eqnarray}
\begin{eqnarray}
  \label{eqnofmotionA}
    A &=& \left(-\frac{\partial }{\partial z}u_k\right)^{\frac32}- 
    \left(-\frac{\partial }{\partial z}u_{k-1}\right)^{\frac32}\\
    B&=&-\frac{\partial }{\partial z}\dot{u}_{k}\sqrt{-\frac{\partial }
    {\partial z}u_{k}} +\frac{\partial }{\partial z}\dot{u}_{k-1}
    \sqrt{-\frac{\partial }{\partial z}u_{k-1}} \label{eqnofmotion2}
\end{eqnarray}
 $A$ and $B$ can be evaluated by Taylor expansion
\begin{eqnarray}
A &=& \left(-\frac{\partial }{\partial _z} u_{k}\right)^\frac{3}{2}- 
\left(-\frac{\partial }{\partial z}u_k + 2R\frac{\partial ^2}{\partial z^2}
u_k\right)^\frac{3}{2} 
= 2R\frac{\partial }{\partial z}\left(-\frac{\partial }{\partial z}
u_k\right)^\frac32  \label{parta}\\
B&=&-\frac{\partial }{\partial z}\dot{u}_k \sqrt{-\frac{\partial }
{\partial z}u_{k}} - \left(-\frac{\partial }{\partial z}\dot{u}_k+
2R\frac{\partial ^2}{\partial z^2}\dot{u}_k\right)\sqrt{-\frac{\partial }
{\partial z}u_k+2R\frac{\partial ^2}{\partial z^2}u_k}\nonumber\\
&\approx& -2R \frac{\partial }{\partial z}\dot{u}_k \frac{\partial }{\partial z}
\sqrt{-\frac{\partial }{\partial z}u_k} -2R \frac{\partial ^2}{\partial z^2}
\dot{u}_k\sqrt{-\frac{\partial }{\partial z}u_k}\\
&=& -2R \frac{\partial }{\partial z}\left(\frac{\partial }{\partial z}\dot{u_k}
\sqrt{-\frac{\partial }{\partial z}u_k}\right)  \label{partb}
\end{eqnarray}
Eqs. (\ref{parta}) and (\ref{partb}) contain only local variables,
therefore we can drop the index $k$ and the equation of motion
(\ref{eqnofmotion1}) in continuum approximation including gravity
reads
\begin{eqnarray}
  \ddot{u}&=&-g-\frac{3\sqrt{2}}{\pi\rho}\left\{\frac{\partial }{\partial z}
\left[\mu\left(-\frac{\partial }{\partial z}u\right)^\frac32+ \alpha 
\left(-\frac{\partial }{\partial z}\dot{u}\sqrt{-\frac{\partial }{\partial z}u}\right)
\right]\right\}\\
&=&-g-\frac{\partial }{\partial z}\left[\kappa\left(-\frac{\partial }{\partial z} 
u\right)^\frac32 - \beta \frac{\partial^2}{\partial t \partial z}u
\sqrt{-\frac{\partial u }{\partial z}}\right]\,.
  \label{motion}
\end{eqnarray}

Eq.~(\ref{motion}) defines the abbreviations $\kappa$ and $\beta$. We
are interested in the critical parameters of driving ($A_0$,
$\omega_0$) when the $N$-th particle loses contact, i.e. when
$u_N>u_{N-1}$. According to the nonlinear interaction of the particles
the motion of all other spheres is not sinusoidal anymore, instead one
finds a superposition of many frequencies. We define the ``response''
$R(\omega_0)$ as the ratio $A_N/A_0$ where $A_N$ is the amplitude of
the $N$-th particle at frequency $\omega_0$ which can be calculated by
convoluting the motion $z_N(t)$ with $\exp(i\omega_0t)$ and $A_0$ is
the amplitude of the driving vibration. Suppose $A_N\omega_N^2/g \ge
1$ the $N$-th particle separates from the $N-1$-st. If we would find
$A_0<A_N$ the critical Froude number $\Gamma_c=A_0\omega_0^2/g$ would
be less than 1. We will show that there is a range for $\omega_0$
where this is the case.

\section{Solution of (\ref{motion}) in the limit of frictionless motion}
The boundary condition for Eq.~(\ref{motion}) is $(\partial u/\partial
z)_{z=L} = 0$. To find the strain of the material without external
forcing (shaking) we consider the limit of no damping ($\alpha=0$).
With
\begin{equation}
x=1-\frac{z}{L},~~~~\tau=\left(\frac{g\kappa^2}{L^5}\right)^\frac{1}{6}t,~~~~  
\Omega=\left(\frac{L^5}{g\kappa^2}\right)^\frac{1}{6}\omega,
\end{equation}
Eq.~(\ref{motion}) turns into 
\begin{equation}
 \frac{\partial^2 u}{\partial \tau^2} = -\gamma^2+\frac{1}{\gamma}
\frac{\partial}{\partial x}\left[\left(\frac{\partial u}{\partial x}
\right)^\frac{3}{2}\right]~~\mbox{with}~~\left(\frac{\partial u}
{\partial x}\right)_{x=0} = 0~,~~~  \gamma=\left(\frac{g^2L^5}{\kappa^2}
\right)^\frac{1}{6}.
\label{undamped}
\end{equation}
Eq.~(\ref{undamped}) is defined for $x\in[0,1]$, its time independent
solution $U(x)$ is
\begin{equation}
  \label{static}
  U(x)=\frac{3}{5}\gamma^2\left(x^\frac{5}{3}-1\right)
\end{equation}
The solution of (\ref{undamped}) can be considered as a superposition
of the static solution (\ref{static}) and a perturbation $w(x,\tau)$.
Inserting $u=U+w$ in (\ref{undamped}) yields
\begin{eqnarray}
  \frac{\partial^2 w}{\partial \tau^2} &=&-\gamma^2+\frac{1}{\gamma}
\frac{\partial}{\partial x}\left[\frac{\partial U}{\partial x} + 
\frac{\partial w}{\partial x}\right]^\frac{3}{2} \nonumber\\
  &=& -\gamma^2+\frac{1}{\gamma}\frac{\partial}{\partial x}\left[\left(
\frac{\partial U}{\partial x}\right)^\frac{3}{2}+\frac{3}{2}\sqrt{
\frac{\partial U}{\partial x}} \frac{\partial w}{\partial x}\right]\nonumber\\
  &=&\frac{3}{2}\frac{\partial}{\partial x}\left[x^\frac{1}{3}
\frac{\partial w}{\partial x}\right]
  \label{wave}
\end{eqnarray}
We are interested in standing wave solutions
$w=T(\tau,\Omega)X(x,\Omega)$ of Eq.~(\ref{wave}) and obtain
\begin{equation}
  \frac{\ddot{T}}{T}=\frac{3}{2X}\frac{\partial }{\partial x}
\left(x^\frac{1}{3}\frac{\partial X}{\partial x}\right)=-\Omega^2
\label{StandingWave}
\end{equation}
with $\Omega$ being a real number. Obviously for $T(\tau,\Omega)$ one
gets $T\sim\exp(i\Omega\tau)$.  The solution of the spatial equation
\begin{equation}
  \label{spatial}
  \frac{3}{2}\frac{\partial }{\partial x}\left(x^\frac{1}{3}
\frac{\partial X}{\partial x}\right)+\Omega^2X=0
\end{equation}
can be found using the Ansatz
\begin{equation}
  \label{ansatz}
  X(x,\Omega)=x^\frac{1}{3}f(y),~~~~~y=\frac{2}{5}\sqrt{6}\Omega x^\frac{5}{6}
\end{equation}
which yields
\begin{equation}
  y^2\frac{\partial ^2f}{\partial y^2}+y\frac{\partial f}{\partial y}+
\left(y^2-\frac{4}{25}\right)f=0\,.
\label{Bessel}
\end{equation}
Eq.~(\ref{Bessel}) is the Bessel equation of order $2/5$. Hence, the
solution of (\ref{spatial}) is
\begin{equation}
  X(x,\Omega)=\left(\frac{6}{25}\right)^\frac{1}{5}\Gamma\left(
\frac{3}{5}\right)\Omega^\frac25 x^\frac{1}{3}J_{-\frac{2}{5}}
\left(\frac{2}{5}\sqrt{6}\Omega x^\frac{5}{6}\right)
\label{Bessel.ort}
\end{equation}
An expression containing $J_{2/5}$ would be a solution too, however,
it does not satisfy the boundary condition of (\ref{undamped}). The
factor has been chosen in order to assure $X(0,\Omega)=1$.

\section{The criterion for the onset of surface motion}

The above defined response $R$ is the ratio $A_N/A_0$ for given
driving frequency $\omega_0$, or $\Omega_0$, respectively. Since the
zeroth particle corresponds to $x=1$ and the $N$-th to $x=0$ we can
write
\begin{equation}
  R^{-1}(\Omega)=\frac{X(1,\Omega)}{X(0,\Omega)} =X(1,\Omega) 
=\left(\frac{6}{25}\right)^\frac{1}{5}\Gamma\left(\frac{3}{5}\right)
\Omega^\frac25J_{-\frac{2}{5}}\left(\frac{2}{5}\sqrt{6}\Omega \right)
  \label{response}
\end{equation}
The response $R$ is an amplification factor, therefore, the value
$g/R(\Omega)$ is the critical acceleration of the driving vibration.
$R$ is larger than 1 for all driving frequencies $\omega$. This means
that for {\em any} driving frequency $\omega_0$ and driving amplitude
$A_0$ the amplitude of the top particle of the column $A_N$ at
frequency $\omega_0$ will be larger than $A_0$. Hence, for $A_N
\omega_0^2/g=1$, i.e. when the $N$-th particle separates from the
$N-1$-st, we find $A_0\omega_0^2/g = \Gamma_c < 1$.

Therefore, we replace the condition $\Gamma\ge 1$ which was supposed
to be the precondition for surface fluidization, convection etc., by
\begin{equation}
  \label{result}
  \frac{A_0\omega_0^2}{g} = \Gamma \ge R^{-1}\left(\omega_0\right).
\end{equation}

The function $R^{-1}(\omega)$ over $\omega$ is drawn in
Fig.~\ref{fig:response} (full line).  The curve reveals pronounced
resonances at Eigenfrequencies $\omega_k$ where $R^{-1}$ becomes
minimal. All experiments on surface fluidization and convection which
can be found in literature have been performed far below the first
resonance. Therefore, of particular interest to practical purposes is
the limit of small frequency $\omega_0$, i.e. below the first
Eigenvalue. The Taylor expansion of $R^{-1}(\Omega)$ for small
$\Omega$ yields
\begin{eqnarray}
  R^{-1}&=&1-\frac{2}{5}\Omega^2+{\cal O}(\Omega^4) = 
1-\frac{2}{5}\left(\frac{L^5}{g\kappa^2}\right)^\frac{1}{3}\omega^2+
{\cal O}(\omega^4)\,.
\label{main}
\end{eqnarray}

Given the container vibrates with frequency $\omega_0$. Then for the
critical amplitude $A_0$ of the vibration when the top particle
separates, i.e. when the material starts to fluidize one finds
\begin{equation}
  A_0=\frac{g}{\omega_0^2}-\frac{2}{5}\left(\frac{L^5}{g\kappa^2}
\right)^\frac{1}{3}\,.
\end{equation}
Surprisingly even for very small frequencies where $R^{-1}\to 1$ one
finds that the critical amplitude is reduced by a constant as compared
with $g/\omega_0^2$.

\section{Numerical solution including damping}

\begin{figure}[tb]
  \centerline{\psfig{figure=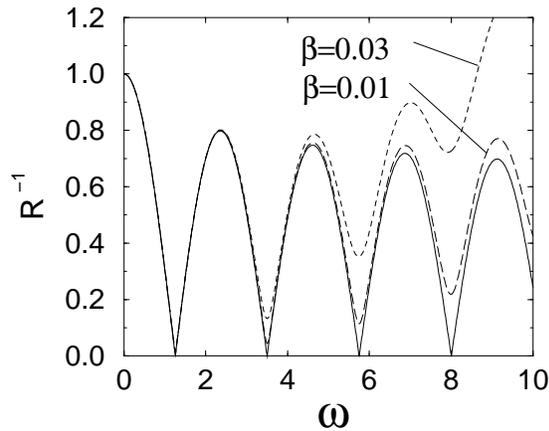,width=7cm}}
\vspace{-0.6cm}
  \caption{The reciprocal response $R^{-1}$ over driving frequency $\omega_0$ 
    without damping, Eq.~(\ref{response}) (full line). The dashed
    lines show numerical results based on integration of
    Eq.~(\ref{netforce}) for damping $\beta=0.01$ and $\beta=0.03$.}
  \label{fig:response}
\end{figure}

Eq.~(\ref{Bessel.ort}) describes the behavior of the column of grains
for the case of purely elastic contact ($\alpha=0$). We have not been
able to solve the full Eq.~(\ref{motion}) including damping in closed
form, therefore we have solved Eq.~(\ref{netforce}) numerically for
different values of damping. The dashed lines in
Fig.~\ref{fig:response} display the reciprocal response $R^{-1}$ over
$\Omega_0$ with fixed amplitude $A=0.1$~mm, elastic constant
$\kappa=2000\,{\rm m}^{7/2}/{\rm s}^2$ and
$L=50$~m. Fig.~\ref{fig:response} 
shows that for small frequency $\Omega_0$ and small damping $\alpha$
the theoretical curve (full line) agrees well with numerical data.

Hence, we believe that at least for small enough damping the results
of the previous section remain correct, i.e. there is a region where
one can find surface fluidization, convection and other macroscopic
effects even if the maximum acceleration lies significantly below $g$.

\section{Conclusion}
We derived an equation of motion for a column of spheres on a
vertically vibrating table, representing a granular material in one
dimensional approximation. It is shown that the sphere on top of the
column $N$ can separate from the $N-1$-st even if the table oscillates
with $\left|A_0\right|\omega^2_0/g = \Gamma < 1$. In agreement with
numerical simulations by Gallas et
al.~\cite{GallasHerrmannSokolowski:1992FLUI} we came to the result
that $\Gamma$ is not the correct criterion for classification of
phenomena which occur in vibrated granular material.

We could show that instead of the widely accepted condition
$A_0\omega_0^2/g>1$ one has to satisfy $A_0\omega_0^2/g>R^{-1}$ where
$R^{-1}$ is a function of $\omega$ which is {\em always} less than
one. Numerical calculations with low damping agree well with analytic
results.

The described result is in contrast with several experimental
investigations where a critical Froude number $\Gamma_c\ge 1$ has been
measured. Whereas the Froude number is certainly the proper criterion
to predict whether a single particle will jump on a vibrating table we
suspect that this number is not suited to be a criterion for surface
fluidization of a column of spheres, and the more not for a three
dimensional granular material.

{\bf Acknowledgements.} The authors wish to thank E.~Cl\'ement,
N.~Gray, H.~J.~Herrmann, H.~M.~Jaeger, S.~Luding, S.~Roux and
L.~Schimansky-Geier for helpful discussions.

\end{document}